\documentclass[preprintnumbers,
amsmath,amssymb,floatfix,10pt,prd,onecolumn,
superscriptaddress,longbibliography,nofootinbib]{revtex4-2}
\usepackage{bm}
\usepackage{amsfonts}
\usepackage{latexsym}
\usepackage{amsmath}
\usepackage{textcomp}
\usepackage{float}
\usepackage{booktabs}
\usepackage{dcolumn}
\usepackage{dsfont}
\usepackage{ragged2e}
\usepackage{epsfig}
\usepackage[dvipsnames]{xcolor}
\usepackage{hyperref}
\hypersetup{           
	colorlinks=true,                
	breaklinks=true,                
	urlcolor= OrangeRed,                
	linkcolor= magenta,                
	bookmarksopen=false,
	filecolor=black,
	citecolor=red,
	linkbordercolor=blue}
\usepackage{graphicx}
\usepackage{overpic}
\usepackage{orcidlink}
\usepackage[T1]{fontenc}

\begin{document}
	\title{Viability of Big Bang Nucleosynthesis in Some Generalized Horizon Entropies}
	
     \author{Kajal Phukan \orcidlink{0009-0003-7732-875X}}
	\email{kajalphukan0@gmail.com}
	\affiliation{%
		Department of Physics, Dibrugarh University, Dibrugarh, Assam, India, 786004}

   \author{Rajdeep Mazumdar \orcidlink{0009-0003-7732-875X}}
	\email[Corresponding author: ]{rajdeepmazumdar377@gmail.com}
	\affiliation{%
		Department of Physics, Dibrugarh University, Dibrugarh, Assam, India, 786004}

  \author{Kalyan Malakar\orcidlink{0009-0002-5134-1553}}%
\email{kalyanmalakar349@gmail.com}
\affiliation{Department of Physics, Dibrugarh University, Dibrugarh, Assam, India, 786004}
\affiliation{Department of Physics, Silapathar College, Dhemaji, Assam, India, 787059}
	
	\author{Kalyan Bhuyan\orcidlink{0000-0002-8896-7691}}%
	\email{kalyanbhuyan@dibru.ac.in}
	\affiliation{%
		Department of Physics, Dibrugarh University, Dibrugarh, Assam, India, 786004}%
	\affiliation{Theoretical Physics Divison, Centre for Atmospheric Studies, Dibrugarh University, Dibrugarh, Assam, India 786004}

	
\begin{abstract}
In this work, we investigate the viability of some cosmological models derived from generalized horizon entropies, using Big Bang Nucleosynthesis (BBN) constraints. By analyzing the deviations in the expansion rate, we derive bounds on the model parameters from freeze-out temperature, helium, and deuterium abundances. Our results show that the freeze-out condition provides the most stringent constraint, while helium and deuterium bounds remain consistent across all models. Although lithium constraints are not satisfied, this discrepancy is attributed to the well-known cosmological lithium problem. Furthermore, the parameter values required for late-time cosmic acceleration are found to lie well within the BBN bounds, demonstrating consistency between early- and late-Universe behavior. These results establish the viability of the considered models within the framework of BBN.
\end{abstract}

	\maketitle

\section{Introduction}
Ever since the formulation of the laws of black hole mechanics by Bardeen
\textit{et al.} and the subsequent understanding of the Bekenstein--Hawking
entropy, a deep and nontrivial connection between gravity and entropy has been
recognized~\cite{bardeen1973,bekenstein1973,hawking1975}. This connection has
been further strengthened by developments such as the holographic principle,
black hole complementarity, gauge/gravity duality, and the firewall puzzle,
all of which strongly suggest that spacetime dynamics, quantum information,
and entropy are intimately related and may play a central role in a comprehensive 
quantum gravity theory. \cite{tHooft1993,susskind1995,maldacena1999,almheiri2013}.\\ Broadly speaking, gravity theories inspired by entanglement and information can be classified into two complementary approaches: holographic gravity and thermodynamic gravity. In the thermodynamic approach, the seminal work of Jacobson demonstrated that the first law of thermodynamics applied to local Rindler horizons can be used to generate Einstein's field equations, suggesting that fundamental thermodynamic principles may give rise to gravitational dynamics.~\cite{jacobson1995}. This idea has since been extended in various directions, leading to alternative derivations of gravitational and
cosmological field equations based on horizon thermodynamics
\cite{padmanabhan2005,verlinde2011}.\\
An important consequence of the gravity \& thermodynamics correspondence is the application of the first rule of thermodynamics to the apparent horizon of Friedmann-Lemaître-Robertson-Walker (FLRW) spacetime to derive cosmological equations. In this framework, the apparent horizon acts as a causal boundary endowed with temperature and entropy, and the Friedmann equations naturally follow from thermodynamic relations \cite{cai2005,akbar2006}. This approach has been successfully applied not only within general relativity but also in a wide class of modified gravity theories, providing a unified thermodynamic interpretation of cosmological
dynamics \cite{sheykhi2018,nojiri2017}.\\
On cosmological scales, observations indicate that the  homogeneous and isotropic nature of the Universe at greater distances and is currently undergoing an accelerated expansion phase. In the standard cosmological model, this acceleration is attributed to a positive cosmological constant, which acts
as dark energy and leads asymptotically to a de Sitter spacetime. An alternative and conceptually appealing description of cosmic acceleration emerges from the thermodynamics of horizons, where modifications to horizon
entropy give rise to effective energy components capable of driving both early- and late-time acceleration. Motivated by the long-range nature of gravity and possible quantum gravity effects, several generalized entropy formalisms have been proposed as extensions of the Bekenstein--Hawking entropy. These include Tsallis, Rényi, Barrow, Kaniadakis, Sharma--Mittal entropies, as well as entropy corrections arising from loop quantum gravity~\cite{tsallis1988,renyi1961,barrow2020,kaniadakis2002,sharma1975,rovelli1996}.
In appropriate limits, these generalized entropies reduce to the standard Bekenstein--Hawking form, while remaining monotonic functions of it. When implemented within horizon thermodynamics, they naturally lead to modified
Friedmann equations and entropy-induced dark energy densities. In this context, Tsallis entropy~\cite{tsallis1988} provides an effective description of non-extensive thermodynamic systems characterized by long-range interactions. Barrow entropy~\cite{barrow2020}, which is closely related to Tsallis entropy, finds natural
applications in quantum gravity scenarios. Rényi entropy is commonly associated with the information-theoretic description of physical systems, while Kaniadakis entropy~\cite{kaniadakis2002} emerges from relativistic statistical mechanics. The Sharma--Mittal entropy~\cite{sharma1975}represents a two-parameter generalization that
unifies the Rényi and Tsallis entropies. Entropy formulations inspired by loop quantum gravity have also been applied
successfully in black hole physics and cosmology. Moreover, recently proposed entropy models have been employed in entropic cosmology and are free from
singular behavior at vanishing Hubble rate $H=0$
\cite{kruglov2025a,kruglov2025b}. 
Other studies of entropies and holographic dark energy models were considered in Refs.~\cite{jahromi2018,ren2021,mejrhit2019,majhi2017,sekhmani2024,noorigashti2025,sadeghi2014,pourhassan2018}.\\
Entropy-modified cosmological models have been widely explored in the context of late-time cosmic acceleration and inflationary dynamics. However, the viability of such models crucially depends on their consistency with early-Universe physics. Big Bang Nucleosynthesis (BBN) \cite{bbn11,bbn12} is one of the most potent and model independent probes of early-Universe physics. The creation of light elements in the early Universe, which took place soon after the Big Bang when the universe was incredibly hot and dense, is known as Big-Bang Nucleosynthesis (BBN) \cite{bbn13}. This epoch spans roughly from about 0.01 seconds to a few hundred seconds (of the order of minutes) after the Big Bang. During this period, the excessive temperatures allowed nuclear reactions to take place, leading to the production of light nuclei such as Helium $^4\mathrm{He}$, Deuterium $\mathrm{D}$, and Lithium $^7\mathrm{Li}$. As the Universe expanded and chilled down, these nuclear reactions progressively became inefficient and eventually stopped, leaving behind the primordial abundances of these elements observed today. Both BBN and the Cosmic Microwave Background (CMB) radiation provide compelling evidence that the early Universe was in a hot and dense state. Following BBN, the Universe underwent several key evolutionary phases, including the radiation-dominated era, the matter-dominated era, and the subsequent formation of large-scale structures.\\
Since even tiny variations in the expansion rate during this period might drastically change the expected abundances of light elements, a cosmological model is considered valid if and only if it meets the strict requirements set by primordial nucleosynthesis. Consequently, Big Bang Nucleosynthesis (BBN) has been extensively
employed to constrain the parameters of modified gravity and non-standard cosmological
models~\cite{bbna1,bbna2,bbna3,bbna4,bbna5}. For instance, Ghoshal and Lambiase \cite{bbn14} studied Tsallis cosmology, finding that the Tsallis parameter is tightly constrained (\(\beta < 2\)) by BBN data, allowing only small deviations from General Relativity. Similarly, Barrow et al. \cite{bbn54} used BBN data to constrain the exponent in Barrow entropy models. Different $f(T, L_m)$ models were examined in Ref.~\cite{s1}, and constraints on their free parameters were derived using observational bounds from the freeze-out temperature with the primordial abundances of light elements. In Ref.~\cite{s2}, Big Bang Nucleosynthesis constraints were studied for dual Kaniadakis cosmology. Furthermore, Ref.~\cite{s3} investigated constraints on Weyl-type $f(Q, T)$ gravity using Big Bang Nucleosynthesis. In particular, modified cosmological scenarios can alter the predicted primordial abundances of light elements. Although the standard BBN model successfully explains the observed abundances of helium and hydrogen, it encounters difficulties in accounting for lithium, giving rise to the well-known cosmological lithium problem. This discrepancy remains an important issue in contemporary cosmology, motivating investigations into whether modified cosmological models can provide a viable resolution.\\
Hence with motivation from the above literature, we employ BBN as a robust probe to constrain some newly introduced modified horizon entropy models arising from horizon thermodynamics, which have been shown to successfully account for the late-time accelerated expansion of the Universe. However, the compatibility of these generalized entropies and the corresponding parametric ranges ensuring their viability with the well-tested physics of the early Universe, particularly Big Bang Nucleosynthesis, remains an important and largely unexplored issue. Hence, in this work, we systematically examine the constraints imposed by BBN on these entropy-driven cosmological models and derive bounds on the parameters for each case as required
for consistency with primordial light-element abundances.

\section{Modified Cosmology Through Generalised Horizon Entropies} \label{s2}
The formulation of modified Friedmann equations using the gravity-thermodynamics conjecture, which may be obtained by using different modified horizon entropies, is briefly reviewed in this section. We consider a Friedmann--Robertson--Walker (FRW) spacetime described by the line element:
\begin{equation}
ds^{2} = -dt^{2} + a^{2}(t)\left(\frac{dr^{2}}{1-k r^{2}} + r^{2} d\Omega^{2}\right),
\label{metric}
\end{equation}
where $a(t)$ is the scale factor and $k=0,+1,-1$ correspond to flat, closed, and open spatial geometries, respectively. We further assume that the universe is filled with conserved perfect fluids consisting of matter and radiation.\\
As stated earlier the first rule of thermodynamics may be applied to the universe horizon as a thermodynamical system separated by a causal boundary, according to the gravity--thermodynamics conjecture~\cite{r1}. The most natural choice for this boundary is the apparent horizon~\cite{r1,r2}, whose radius is given by
\begin{equation}
r_A = \frac{1}{\sqrt{H^{2} + \frac{k}{a^{2}}}},
\label{apparent_horizon}
\end{equation}
where $H = \dot{a}/a$ is the Hubble parameter and an overdot denotes differentiation with respect to cosmic time. The temperature associated with the apparent horizon is taken to be analogous to the black hole temperature, namely~\cite{r3}
\begin{equation}
T_h = \frac{1}{2\pi r_A}.
\label{horizon_temperature}
\end{equation} 
Similarly, the Bekenstein-Hawking area law defines the entropy of the apparent horizon as:
\begin{equation}
S_h = \frac{A}{4G},
\label{bh_entropy}
\end{equation}
where $A = 4\pi r_A^{2}$ is the horizon area and $G$ is the gravitational constant (we work in units $\hbar = k_B = c = 1$). The energy flux crossing the apparent horizon during an infinitesimal time interval $dt$ is
\begin{equation}
\delta Q = -dE = A (\rho_m + p_m + \rho_r + p_r) H r_A \, dt,
\label{energy_flux}
\end{equation}
where $\rho_i$ and $p_i$ denote the energy densities and pressures of matter and radiation, respectively. Employing the first law of thermodynamics,
\begin{equation}
-dE = T_h \, dS_h,
\label{first_law}
\end{equation}
together with Eqs.~\eqref{horizon_temperature} and \eqref{bh_entropy}, one obtains the standard Friedmann equations~\cite{cai2005}
\begin{equation}
-4\pi G (\rho_m + p_m + \rho_r + p_r)
= \dot{H} - \frac{k}{a^{2}},
\label{friedmann_1}
\end{equation}
\begin{equation}
\frac{8\pi G}{3} (\rho_m + \rho_r)
= H^{2} + \frac{k}{a^{2}} - \frac{\Lambda}{3},
\label{friedmann_2}
\end{equation}
where the cosmological constant $\Lambda$ is an integration constant. It is noteworthy that the equilibrium assumption, which states that the temperature of the cosmic fluids is equal to that of the apparent horizon~\cite{r1,r2}, is used to prevent non-equilibrium thermodynamics.\\
The gravity--thermodynamics correspondence has been extensively applied to various modified theories of gravity by appropriately modifying the entropy--area relation~\cite{k1,t,b,k3,x,ref90}. Here, using different modified horizon entropies instead of Eq.~\eqref{bh_entropy}, one finally obtains the modified Friedmann equations corresponding to the respective modified entropy theory. The modifications induced by generalized horizon entropies can be interpreted as an effective dark energy sector. Once the modified Friedmann equations are derived from a modefied entropy theory, they can always be rewritten in the standard form:
\begin{equation}
H^{2} + \frac{k}{a^{2}} = \frac{8\pi G}{3} \left( \rho_m + \rho_r + \rho_{\rm de} \right),
\end{equation}
\begin{equation}
\dot{H} - \frac{k}{a^{2}} = -4\pi G \left( \rho_m + p_m + \rho_r + p_r + \rho_{\rm de} + p_{\rm de} \right),
\end{equation}
where $\rho_{\rm de}$ and $p_{\rm de}$ effectively encode all deviations from standard Einstein gravity arising from the modified entropy. Hence, for any modified horizon entropy, the corresponding effective dark energy density and pressure can be directly identified by comparison with the above equations, allowing a unified effective-fluid description of the resulting cosmological dynamics.

\section{Big Bang Nucleosynthesis Constraints} \label{s3}
In this section, we analyze the constraints arising from Big Bang Nucleosynthesis (BBN) within the framework of the modified cosmological model. We first derive the deviation in the freeze-out temperature and subsequently study the implications for the primordial abundances of light elements.

\subsection{Freeze-out Temperature and BBN Bound}
Given that BBN takes place in the radiation-dominated period \cite{bbn54,70}, the first Friedmann equation in the context of standard general relativity (GR) can be approximated as:
\begin{equation}
H^{2} \simeq \frac{\rho_{r}}{3M_{p}^{2}} \equiv H_{\rm GR}^{2},
\label{bbn_fried}
\end{equation}
where $\rho_{r}$ represents the energy density of relativistic particles that make up the universe and $M_{p}=(8\pi G)^{-1/2}$ is the reduced Planck mass. The radiation energy density is given by:
\begin{equation}
\rho_{r} = \frac{\pi^{2}}{30} g_{*} T^{4},
\label{rho_r}
\end{equation}
where $T$ is the temperature and $g_{*}=g(T)$ represents the effective number of relativistic degrees of freedom, typically approximated as $g_{*}\simeq 10$ during the BBN epoch. Substituting Eqs.~\eqref{bbn_fried} and \eqref{rho_r}, the Hubble parameter as a function of temperature becomes
\begin{equation}
H(T) = \left(\frac{4\pi^{3}g_{*}}{45}\right)^{1/2}\frac{T^{2}}{M_{p}},
\label{H_T}
\end{equation}
where $M_{p}=\sqrt{8\pi}\,M_{P}$ denotes the Planck mass. Owing to radiation conservation, the scale factor evolves as $a(t)\propto \sqrt{t}$, where $t$ is the cosmic time. Consequently, the Hubble parameter can be written as $H=1/(2t)$, which leads to the temperature–time relation
\begin{equation}
\frac{1}{t} = \left(\frac{16\pi^{3}g_{*}}{45}\right)^{1/2}\frac{T^{2}}{M_{p}},
\end{equation}
or equivalently $T(t)\propto (t/{\rm s})^{-1/2}\,{\rm MeV}$. During BBN, neutrons are produced via proton–neutron conversion processes \cite{70,71}
\begin{align}
\Gamma_{pn}(T) = {} & (n+\nu_{e}\rightarrow p+e^{-}) 
+ (n+e^{+}\rightarrow p+\bar{\nu}_{e}) \nonumber \\
& + (n\rightarrow p+e^{-}+\bar{\nu}_{e}),
\label{pn_rate}
\end{align}
along with the inverse processes $\Gamma_{np}(T)$. The total interaction rate is thus given by
\begin{equation}
\Gamma_{\rm tot}(T)=\Gamma_{pn}(T)+\Gamma_{np}(T).
\end{equation}
Assuming all particles are at the same temperature and sufficiently relativistic, the Boltzmann distribution can be employed instead of the Fermi–Dirac distribution. The overall interaction rate can be expressed as follows, ignoring the electron mass in relation to electron and neutrino energies: \cite{72,73}
\begin{equation}
\Gamma_{\rm tot}(T) = 8\left(12T^{2}+6QT+Q^{2}\right)AT^{3},
\label{tot_rate}
\end{equation}
where $Q=m_{n}-m_{p}=1.29\times10^{-3}\,{\rm GeV}$ is the neutron–proton mass difference and $A=1.02\times10^{-11}\,{\rm GeV}^{-4}$. The primordial helium mass fraction is given by
\begin{equation}
Y_{p}=\frac{2\lambda\,x(T_{f})}{1+x(T_{f})},
\end{equation}
where $\lambda=\exp[(T_{f}-T_{n})/\tau]$, $T_{f}$ denotes the freeze-out temperature of weak interactions, $T_{n}$ represents the onset of nucleosynthesis, $x(T_{f})=\exp(-Q/T_{f})$ is the neutron-to-proton equilibrium ratio, and $\tau=880.3\pm1.1\,{\rm s}$ is the neutron mean lifetime \cite{74}. The function $\lambda(T_{f})$ accounts for neutron decay in the interval $[T_{f},T_{n}]$. The freeze-out temperature is obtained by comparing the interaction time scale $\Gamma_{\rm tot}^{-1}$ with the expansion time scale $H^{-1}$. Thermal equilibrium is maintained when $H^{-1}\gg\Gamma_{\rm tot}^{-1}$, whereas decoupling occurs when $H^{-1}\ll\Gamma_{\rm tot}^{-1}$. The freeze-out condition is therefore given by
\begin{equation}
H(T_{f})=\Gamma_{\rm tot}(T_{f}) \simeq c_{q} T_{f}^{5},
\end{equation}
where $c_{q}\equiv 96A \simeq 9.8\times10^{-10}\,{\rm GeV}^{-4}$ \cite{bbn54,72,73}. Using Eq.~\eqref{H_T}, the freeze-out temperature becomes
\begin{equation}
T_{f}=\left(\frac{4\pi^{3}g_{*}}{45\,c_{q}^{2}M_{p}^{2}}\right)^{1/6}.
\label{Tf_standard}
\end{equation}
In modified cosmological models, the Hubble parameter deviates from its GR counterpart, leading to a shift in the freeze-out temperature $T_{f}\rightarrow T_{f}+\Delta T_{f}$. Consequently, the helium mass fraction acquires a correction given by
\begin{equation}
\Delta Y_{p}
=Y_{p}\left[\frac{1-Y_{p}}{2\lambda}\ln\left(\frac{2\lambda}{Y_{p}}-1\right)
-\frac{2T_{f}}{\tau}\right]\frac{\Delta T_{f}}{T_{f}},
\label{deltaY}
\end{equation}
where we have taken $T(T_{n})=0$, since $T_{n}$ is fixed by the deuterium binding energy \cite{72,73,75}. Observationally, the primordial helium abundance is constrained as \cite{76}
\begin{equation}
Y_{p}=0.2476, \qquad |\Delta Y_{p}|<10^{-4}.
\label{Yp_obs}
\end{equation}
In modified gravity theories or in the presence of an additional dark energy component, the Friedmann equation can be written as
\begin{equation}
3M_{p}^{2}H^{2}=\rho_{m}+\rho_{r}+\rho_{\rm DE}.
\end{equation}
During the BBN era, the matter contribution is negligible, yielding
\begin{equation}
H = H_{\rm GR}\left(1+\frac{\rho_{\rm DE}}{\rho_{r}}\right)^{1/2},
\label{H_mod}
\end{equation}
where $H_{\rm GR}$ denotes the expansion rate in standard cosmology. For $\rho_{\rm DE}\ll\rho_{r}$, this reduces to
\begin{equation}
H \simeq H_{\rm GR}\left(1+\frac{1}{2}\frac{\rho_{\rm DE}}{\rho_{r}}\right).
\end{equation}
This deviation induces a corresponding shift in the freeze-out temperature. Using the relation $H_{\rm GR}\simeq c_{q}T_{f}^{5}$ along with Eq.~\eqref{Tf_standard}, one finds
\begin{equation}
\frac{\Delta T_{f}}{T_{f}} \simeq
\frac{\rho_{\rm DE}}{\rho_{r}}
\frac{H_{\rm GR}}{10\,c_{q}T_{f}^{5}}.
\label{deltaTf}
\end{equation}
Finally, the theoretically predicted deviation must satisfy the observational bound
\begin{equation}
|\frac{\Delta T_{f}}{T_{f}}| < 4.7\times10^{-4},
\label{Tf_bound}
\end{equation}
which follows from the observational constraints on the primordial helium mass fraction given in Eq.~\eqref{Yp_obs}.\\
For the following sections to study primordial abundances of light elements in more details, we introduce the dimensionless parameter
\begin{equation}
Z \equiv \frac{H}{H_{\mathrm{GR}}} = \left(1+\frac{\rho_{\rm DE}}{\rho_{r}}\right)^{1/2},
\label{z}
\end{equation}
which measures the ratio of the expansion rate in the modified model to that in General Relativity (GR).
\subsection{Helium \texorpdfstring{$^4\mathrm{He}$}{4He} Abundance}

The production of $^4\mathrm{He}$ proceeds through a sequence of nuclear reactions beginning with the formation of deuterium \cite{27}:
\begin{align}
n + p &\rightarrow {}^2\mathrm{H} + \gamma, \\
{}^2\mathrm{H} + {}^2\mathrm{H} &\rightarrow {}^3\mathrm{He} + n, \\
{}^2\mathrm{H} + {}^2\mathrm{H} &\rightarrow {}^3\mathrm{H} + p,
\end{align}
followed by
\begin{align}
{}^2\mathrm{H} + {}^3\mathrm{H} &\rightarrow {}^4\mathrm{He} + n, \\
{}^2\mathrm{H} + {}^3\mathrm{He} &\rightarrow {}^4\mathrm{He} + p.
\end{align}

The numerical best-fit expression for the primordial helium abundance is given by\cite{31}:
\begin{equation}
Y_p = 0.2485 \pm 0.0006 + 0.0016 \left[(\eta_{10} - 6) + 100(Z - 1)\right],
\end{equation}
where the baryon-to-photon ratio is defined as\cite{4}:
\begin{equation}
\eta_{10} \equiv 10^{10} \frac{n_B}{n_\gamma} \simeq 6.
\end{equation}
Using the observational constraint $Y_p = 0.2449 \pm 0.0040$\cite{29}, we obtain the following bound on $Z$:
\begin{equation}
Z = 1.0475 \pm 0.105.
\label{a1}
\end{equation}

\subsection{Deuterium \texorpdfstring{$^2\mathrm{H}$}{2H} Abundance}

Deuterium is primarily produced through the reaction\cite{27}:
\begin{equation}
n + p \rightarrow {}^2\mathrm{H} + \gamma.
\end{equation}
The corresponding best-fit formula for its abundance is\cite{4}:
\begin{equation}
Y_{D_p} = 2.6(1 \pm 0.06) \left(\frac{6}{\eta_{10} - 6(Z - 1)}\right)^{1.6}.
\end{equation}
By comparing with the observational value $Y_{D_p} = 2.55 \pm 0.03$ \cite{29}, we obtain
\begin{equation}
Z = 1.062 \pm 0.444,
\label{a2}
\end{equation}
which is consistent with the constraint obtained from helium abundance.

\subsection{Lithium \texorpdfstring{$^7\mathrm{Li}$}{7Li} Abundance}

The production of $^7\mathrm{Li}$ occurs via the nuclear processes \cite{4}:
\begin{align}
{}^3\mathrm{H} + {}^4\mathrm{He} &\rightarrow {}^7\mathrm{Li} + \gamma, \\
{}^3\mathrm{He} + {}^4\mathrm{He} &\rightarrow {}^7\mathrm{Be} + \gamma, \\
{}^7\mathrm{Be} + n &\rightarrow {}^7\mathrm{Li} + p.
\end{align}

The best-fit expression for lithium abundance is\cite{4,31}:
\begin{equation}
Y_{\mathrm{Li}} = 4.82(1 \pm 0.1) \left(\frac{\eta_{10} - 3(Z - 1)}{6}\right)^2.
\end{equation}
Using the observational constraint $Y_{\mathrm{Li}} = 1.6 \pm 0.3$ \cite{29}, we obtain
\begin{equation}
Z = 1.960025 \pm 0.076675.
\label{a3}
\end{equation}
It is evident that the constraint on $Z$ obtained from lithium abundance does not overlap with those derived from deuterium and helium. This discrepancy reflects the well-known cosmological lithium problem \cite{57}, where the predicted abundance of $^7\mathrm{Li}$ exceeds the observed value by a factor of approximately $2.4$--$4.3$. While the standard BBN scenario successfully explains the abundances of deuterium and helium, it fails to account for lithium, and this issue persists even in many modified cosmological models. Therefore, consistency with deuterium and helium abundances is generally considered sufficient for BBN viability, whereas lithium remains an open problem in modern cosmology.

\section{BBN Constraints on the Modified Entropies}
In this section based on the constraints and the framework as  discussed in Sec.(\ref{s2}) and Sec.(\ref{s3}), we investigate the BBN constraints on some recently introduced modified entropy models. As, stated previously in this work we consider three different cases of modified entropy models. The respective models and the BBN constraints on them are as shown below. 
\subsection{Model I}
The first model we consider was recently proposed by Kruglov~\cite{k1}, who introduced a modified apparent horizon entropy of the form:
\begin{equation}
S_{K1} = \frac{S_{\rm BH}}{1 + \gamma S_{\rm BH}},
\end{equation}
where $S_{\rm BH}$ denotes the Bekenstein--Hawking entropy and $\gamma$ is a positive constant encoding deviations from the standard form. This entropy increases positively and monotonically with $S_{\rm BH}$ and disappears at $S_{\rm BH}=0$. In the limit $\gamma = 0$, the standard Bekenstein--Hawking entropy is smoothly recovered. Employing this entropy within the gravity--thermodynamics correspondence leads to modified Friedmann equations featuring an effective dynamical cosmological term, which can account the late time accelerated expansion of the universe well discussed in Ref.~\cite{k1}.\\
For this modified entropy one finds the density of dark
energy for a flat space-time as:
\begin{equation}
\rho_{de} = \frac{3 b}{8 \pi G} \left(\frac{b}{b+H^2}+2 \log \left(\frac{b+H^2}{b}\right)\right)
\label{E1}
\end{equation}
where, $b = \frac{\pi  \gamma}{G} $ and $G = \frac{1}{8 \pi M_p^{2}}$.  In Eq.(\ref{E1}) we may approximate the Hubble parameter by its general relativistic expression, $H \approx H_{\mathrm{GR}}$, and $\rho_r$ by it's standard expression, using which in Eqs.(\ref{deltaTf}) \& (\ref{z}) gives us:
\begin{equation}
\frac{\Delta T_{f}}{T_{f}} = \frac{3 b M_p \left(\frac{b}{b+\frac{\pi ^2 g_{*} T_f^4}{90 M_p^2}}+2 \log \left(\frac{\pi ^2 g_{*} T_f^4}{90 b M_p^2}+1\right)\right)}{\sqrt{10} \pi  \sqrt{g_{*}} C_q T_f^7}
\label{e1}
\end{equation}
\begin{equation}
Z = \frac{45 b M_p^2 \left(\frac{b}{b+\frac{\pi ^2 g_{*} T^4}{90 M_p^2}}+2 \log \left(\frac{\pi ^2 g_{*} T^4}{90 b M_p^2}+1\right)\right)}{\pi ^2 g T^4}+1
\label{e2}
\end{equation}
Now, using Eq. (\ref{e1}) with Eq. (\ref{Tf_bound}), and Eq. (\ref{e2}) with Eqs. (\ref{a1}), (\ref{a2}), and (\ref{a3}, the viability  constraint on the model parameter $b$ is inferred. 
\begin{figure*}[htb]
\centerline{
\includegraphics[width=1.01\textwidth]{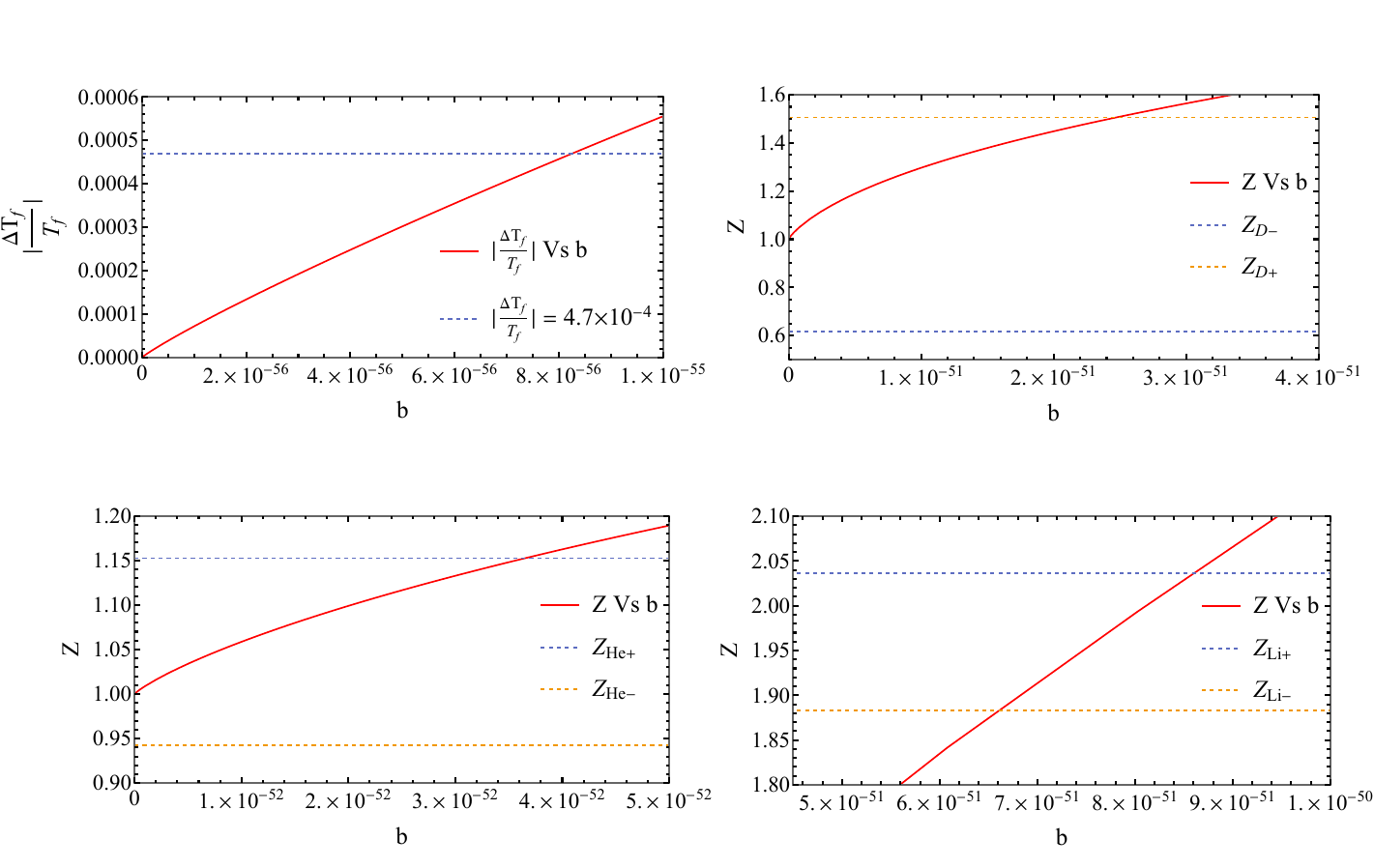}}
\caption{plot of $|\frac{\Delta T_{f}}{T_{f}}|$ \& $Z$ Vs $b$ for constraints on model I. With $T=1MeV$.}
\label{f1}
\end{figure*}
As shown in Fig.~(\ref{f1}) and Table~\ref{t1}, the constraints obtained from freeze-out temperature, helium, and deuterium abundances are mutually consistent and impose an upper bound on the model parameter as
\begin{equation}
0 < b < 8.26 \times 10^{-56}.
\end{equation}
This indicates that the freeze-out condition provides the most stringent constraint on the model parameter.

On the other hand, the constraint obtained from lithium abundance,
\begin{equation}
6.6 \times 10^{-51} < b < 8.6 \times 10^{-51},
\end{equation}
does not overlap with the above range. This inconsistency reflects the well-known cosmological lithium problem, which persists even within the standard $\Lambda$CDM framework. Therefore, the model can still be considered viable from the perspective of BBN, as it successfully satisfies the constraints from helium and deuterium abundances.

Ref.~\cite{k1} suggests that for $b \simeq \dfrac{H_0^{2}}{5.819} \sim 10^{-84}\,\mathrm{GeV}^2$, the model can successfully account for the late-time accelerated expansion of the Universe. Notably, this value of $b$ is also consistent with the bounds imposed by Big Bang Nucleosynthesis (BBN) in the present analysis, indicating that the model remains viable across both early- and late-time cosmological regimes.

‌\subsection{Model II}
The second model we consider is characterized by a modified apparent horizon entropy of the form~\cite{kruglov2025b}:
\begin{equation}
S_{K3} = \frac{S_{\rm BH}}{1 + \lambda S_{\rm BH}^2}
\end{equation}
where $S_{\rm BH}$ again denotes the Bekenstein--Hawking entropy and $\lambda$ is a positive constant that quantifies deviations from the standard form. Just like the previous model this entropy is also a positive and monotonically increasing function of $S_{\rm BH}$ and vanishes when $S_{\rm BH}=0$. Similarly, in the limit $\lambda = 0$, the standard Bekenstein--Hawking entropy is smoothly recovered. According to Ref.~\cite{kruglov2025b}, cosmology based on the modified Friedmann equations derived from it might be very useful for explaining inflation and the acceleration of the late time universe.
For this modified entropy one finds the density of dark
energy for a flat space-time as:
\begin{equation}
\rho_{de} = \frac{3}{8\pi G}
\left[
2\sqrt{b}\,\arctan\!\left(\frac{H^2}{\sqrt{b}}\right)
- \frac{b\,H^2}{b + H^4}
\right],
\label{E2}
\end{equation}
where, $G = \frac{1}{8 \pi M_p^{2}}$ and $b = \frac{\lambda \pi}{G^2} $. Just like in the above case using Eq.(\ref{E2}) in Eqs.(\ref{deltaTf}) \& (\ref{z}) gives us:
\begin{equation}
\frac{\Delta T_{f}}{T_{f}} = \frac{6 \sqrt{b} M_p^2 \tan ^{-1}\left(\frac{\pi ^2 g_{*} T_f^4}{90 \sqrt{b} M_p^2}\right)-\frac{270 \pi ^2 b g_{*} T_f^4 M_p^4}{8100 b M_p^4+\pi ^4 g_{*}^2 T_f^8}}{\sqrt{10} \pi  C_q \sqrt{g_{*}} T_f^7 M_p}
\label{e12}
\end{equation}
\begin{equation}
Z = \frac{15 \left(6 \sqrt{b} M_p^2 \tan ^{-1}\left(\frac{\pi ^2 g_{*} T^4}{90 \sqrt{b} M_p^2}\right)-\frac{270 \pi ^2 b g_{*} T^4 M_p^4}{8100 b M_p^4+\pi ^4 g_{*}^2 T^8}\right)}{\pi ^2 g_{*} T^4}+1
\label{e22}
\end{equation}
Now, using Eq. (\ref{e12}) with Eq. (\ref{Tf_bound}), and Eq. (\ref{e22}) with Eqs. (\ref{a1}), (\ref{a2}), and (\ref{a3}, the viability  constraint on the model parameter $b$ is inferred. 
\begin{figure*}[htb]
\centerline{
\includegraphics[width=1.01\textwidth]{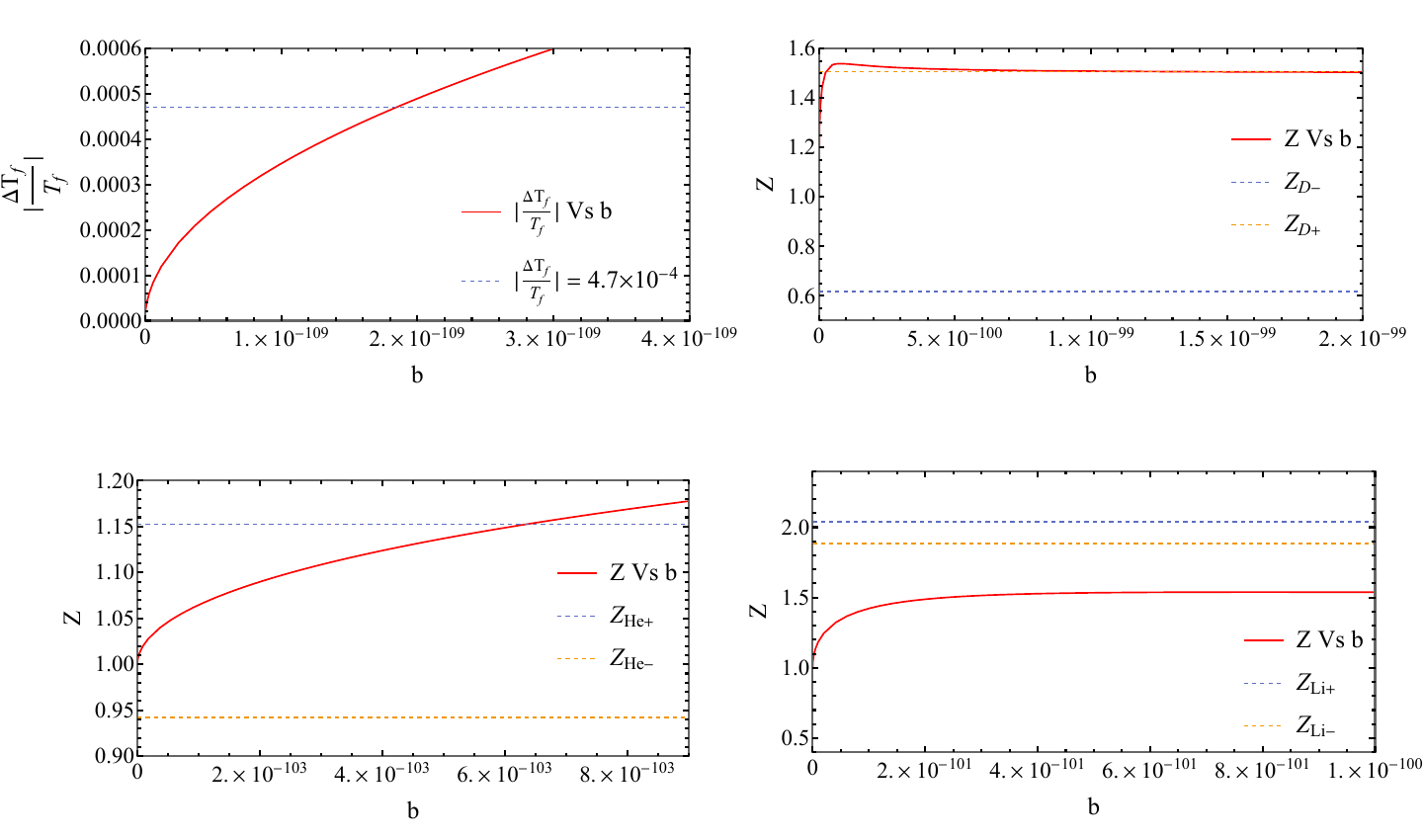}}
\caption{plot of $|\frac{\Delta T_{f}}{T_{f}}|$ \& $Z$ Vs $b$ for constraints on model II. With $T=1MeV$.}
\label{f2}
\end{figure*}
As shown in Fig.~(\ref{f2}) and Table~\ref{t1} for the present model, the constraints obtained from freeze-out temperature, helium, and deuterium abundances are consistent and lead to the bound:
\begin{equation}
0 < b < 1.84 \times 10^{-109}.
\end{equation}
It is evident that the freeze-out condition again provides the most stringent constraint on the model parameter. Although the deuterium constraint admits an additional branch $b > 1.41 \times 10^{-99}$, this range does not overlap with the bounds obtained from freeze-out and helium, and is therefore not physically relevant. Furthermore, no consistent solution is obtained from the lithium abundance, reflecting the persistence of the cosmological lithium problem. Thus, the model remains viable from the perspective of BBN, provided the parameter $b$ satisfies the above bound, ensuring consistency with helium and deuterium observations.\\
Furthermore, Ref.~\cite{kruglov2025b} suggests that for the given model $b \simeq \left(\frac{H}{1.23}\right)^4$, hence during the late-time the model parameter becomes $b \simeq \left(\frac{H_0}{1.23}\right)^4 \sim 10^{-168}\,\mathrm{GeV}^4$. Which lies many orders of magnitude below the upper bound obtained from BBN constraints, $b < 10^{-109}$. This demonstrates that the model is naturally consistent with early-Universe nucleosynthesis requirements while simultaneously accounting for the late-time cosmic expansion.

\subsection{Model III}
The third model we consider is characterized by a modified apparent horizon entropy of the form~\cite{x}:
\begin{equation}
S_{K3} = \frac{S_{\rm BH}}{{\left(1 + \sigma S_{\rm BH}\right)}^2}
\end{equation}
where $S_{\rm BH}$ again denotes the Bekenstein--Hawking entropy and $\sigma$ is a positive constant that quantifies deviations from the standard form. Just like the previous two models this entropy is also a positive and monotonically increasing function of $S_{\rm BH}$ and vanishes when $S_{\rm BH}=0$. Similarly, in the limit $\sigma = 0$, the standard Bekenstein--Hawking entropy is smoothly recovered. Cosmology based on the modified Friedmann equations derived from it can also be of significant relevance for a description of inflation and late-time universe acceleration, as detailed in Ref.~\cite{x}. For this modified entropy one finds the density of dark
energy for a flat space-time as:
\begin{equation}
\rho_{de} = \frac{3b}{8 \pi G} \left(\frac{b \left(4 b+5 H^2\right)}{\left(b+H^2\right)^2}+4 \log \left(\frac{b+H^2}{b}\right)\right),
\label{E1c}
\end{equation}
where, $b = \frac{\pi  \sigma}{G} $ and $G = \frac{1}{8 \pi M_p^{2}}$. Just like in the above case using Eq.(\ref{E1c}) in Eqs.(\ref{deltaTf}) \& (\ref{z}) gives us:
\begin{equation}
\frac{\Delta T_{f}}{T_{f}} = \frac{3 b M_p \left(\frac{450 b M_p^2 \left(72 b M_p^2+\pi ^2 g_{*} T_f^4\right)}{\left(90 b M_p^2+\pi ^2 g_{*} T_f^4\right){}^2}+4 \log \left(\frac{\pi ^2 g_{*} T_f^4}{90 b M_p^2}+1\right)\right)}{\sqrt{10} \pi  C_q \sqrt{g_{*}} T_f^7}
\label{e13}
\end{equation}
\begin{equation}
Z = \frac{45 b M_p^2 \left(\frac{450 b M_p^2 \left(72 b M_p^2+\pi ^2 g_{*} T^4\right)}{\left(90 b M_p^2+\pi ^2 g_{*} T^4\right){}^2}+4 \log \left(\frac{\pi ^2 g_{*} T^4}{90 b M_p^2}+1\right)\right)}{\pi ^2 g_{*} T^4}+1
\label{e23}
\end{equation}
Using Eq. (\ref{e13}) with Eq. (\ref{Tf_bound}), and Eq. (\ref{e23}) with Eqs. (\ref{a1}), (\ref{a2}), and (\ref{a3}, the viability  constraint on the model parameter $b$ is inferred. 
\begin{figure*}[htb]
\centerline{
\includegraphics[width=1.01\textwidth]{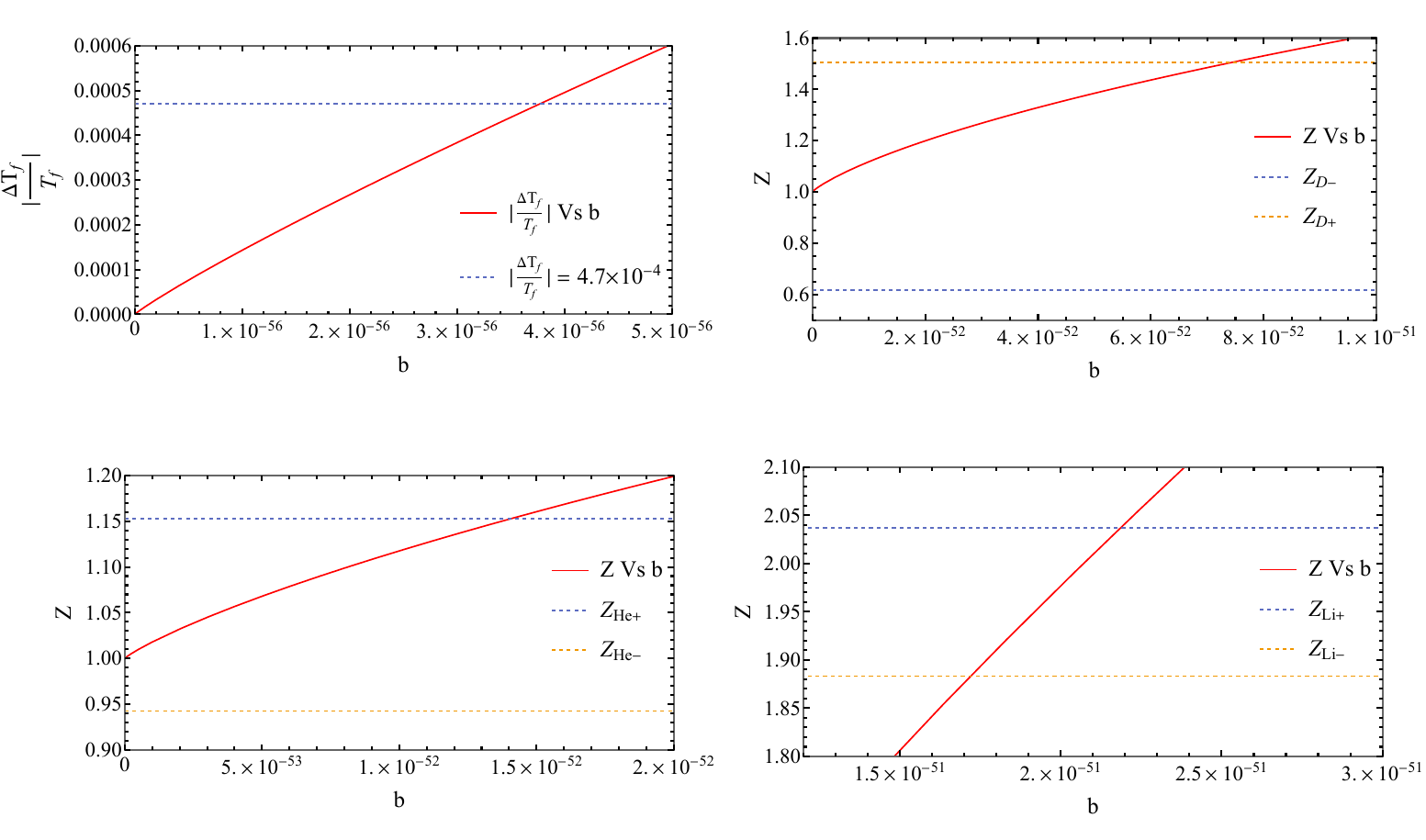}}
\caption{plot of $|\frac{\Delta T_{f}}{T_{f}}|$ \& $Z$ Vs $b$ for constraints on model III. With $T=1MeV$.}
\label{f3}
\end{figure*}
As shown in Fig.~(\ref{f3}) and Table~\ref{t1} for the present model,For this model, the constraints obtained from freeze-out temperature, helium, and deuterium abundances are mutually consistent and lead to the bound
\begin{equation}
0 < b < 3.77 \times 10^{-56}.
\end{equation}
It is evident that the freeze-out condition here also provides the most stringent constraint on the model parameter. The lithium abundance, however, yields the constraint
\begin{equation}
1.72 \times 10^{-51} < b < 2.18 \times 10^{-51},
\end{equation}
which does not overlap with the above range, reflecting towards the lithium problem that persist for this model also. Therefore, the model can be considered viable from the perspective of BBN, as it satisfies the constraints from helium and deuterium abundances while the lithium inconsistency remains an open issue.\\
Furthermore, Ref.~\cite{x} suggests that for $b \simeq \dfrac{H_0^{2}}{17.4} \sim 10^{-84}\,\mathrm{GeV}^2$ the model can successfully account for the late-time accelerated expansion of the Universe. Notably, this value of $b$ is also consistent with the bounds imposed by Big Bang Nucleosynthesis (BBN) in the present analysis, indicating that the model remains viable across both early- and late-time cosmological regimes.
\subsection{Model IV}
\begin{figure*}[htb]
\centerline{
\includegraphics[width=1.01\textwidth]{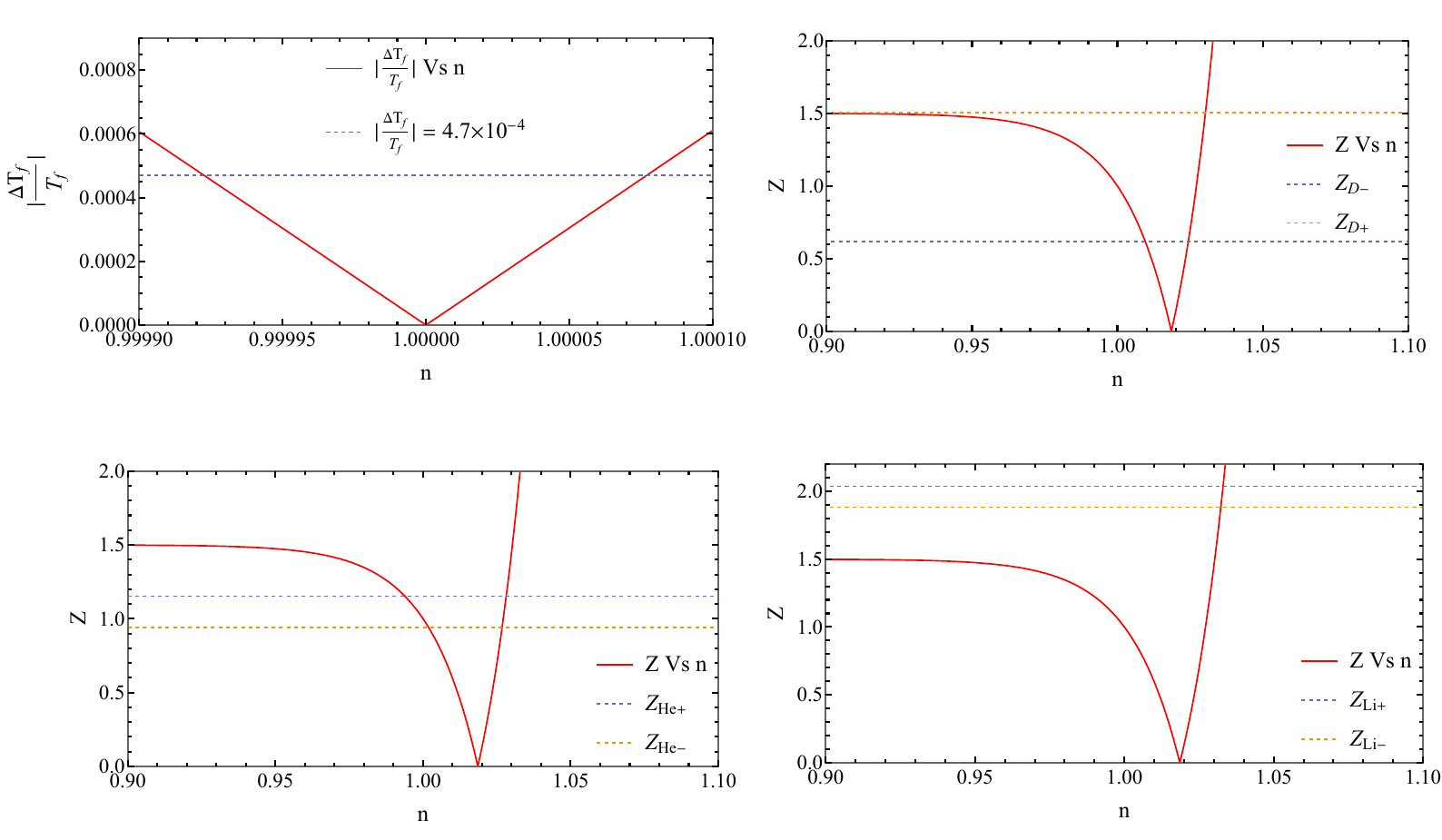}}
\caption{plot of $|\frac{\Delta T_{f}}{T_{f}}|$ \& $Z$ Vs $\gamma$ for constraints on model IV. With $T=1MeV$.}
\label{f4}
\end{figure*}

As the fourth model we consider a generalized mass-to-horizon entropy model proposed in the literature \cite{ref90}. This model introduces a two-parameter extension of the Bekenstein–Hawking entropy, arising from a modified mass-to-horizon relation. The corresponding entropy is given by
\begin{equation}
S_n = \gamma \frac{2n}{n+1} r_A^{\,n-1} S_{BH},
\end{equation}
where $S_{BH}$ denotes the standard Bekenstein–Hawking entropy, $r_A$ represents the radius of the apparent horizon, and $\gamma$ and $n$ are positive valued free parameters characterizing deviations from the standard case. It is worth noting that for $\gamma = 1$ and $n = 1$, the standard Bekenstein–Hawking entropy and the usual mass-to-horizon relation are recovered.\\
For this modified entropy one finds the density of dark
energy for a flat space-time as \cite{ref90,xx}:
\begin{equation}
\rho_{de} = \frac{3}{8 \pi  G} \left(H^2-\frac{2 \gamma  n H^{3-n}}{3-n}\right),
\label{E1d}
\end{equation}
where, $G = \frac{1}{8 \pi M_p^{2}}$. Using Eq.(\ref{E1d}) in Eqs.(\ref{deltaTf}) \& (\ref{z}) gives us:
\begin{equation}
\frac{\Delta T_{f}}{T_{f}} = \frac{\left(\frac{\sqrt{g_*} T_f^2}{M_p}\right){}^{-n} \left(3 \pi  \sqrt{10} g_* (n-3) T_f^4 \left(\frac{\sqrt{g_*} T_f^2}{M_p}\right){}^{n-1}+\gamma  g_* 2^{\frac{n}{2}+1} 3^n 5^{n/2} n \pi ^{2-n} T_f^4\right)}{900 (n-3) C_q \left(T_f\right)_f^5 M_p^2}
\label{e14}
\end{equation}
\begin{equation}
Z = \frac{\gamma  g_* 3^{n-1} 10^{\frac{n-1}{2}} n \pi ^{1-n} T^4 \left(\frac{\sqrt{g_*} T^2}{M_p}\right){}^{-n-1}}{(n-3) M_p^2}+\frac{3}{2}
\label{e24}
\end{equation}
Using Eq. (\ref{e14}) with Eq. (\ref{Tf_bound}), and Eq. (\ref{e24}) with Eqs. (\ref{a1}), (\ref{a2}), and (\ref{a3}, the viability  constraint on the model parameters can be inferred. According to Refs. (\cite{ref90,xx}), $\gamma$ emerges as an overall multiplicative factor and simply modifies the entropy-induced terms' amplitude without changing their functional dependence,  so mostly $\gamma$ is taken as unity through out different studies. Furthermore, any deviation from unity is anticipated to be insignificant or, at most, sub-dominant in comparison to the effects brought about by changes in the exponent n. But,the parameter $n$ can significantly alter the cosmological dynamics of the late time universe. In Ref.~\cite{xy}, a combined analysis of Type Ia Supernovae, Cosmic Chronometers, Baryon Acoustic Oscillations(including the DESI DR2 release), and SH0ES datasets constrains the model parameter to $n = 0.945 \pm 0.070$. Similarly, in Ref.~\cite{xx} baryogenesis constraints on parameter n gives $0.98 \lesssim n < 1$. And, so as shown in Fig. (\ref{f4}) and Table (\ref{t1}) taking $\gamma = 1$ for this model, the constraints obtained from freeze-out temperature, helium, and deuterium abundances are mutually consistent and lead to a narrow allowed range for the model parameter $n$, given by
\begin{equation}
0.999923 < n < 1.00008
\end{equation}
Here, also the freeze-out condition provides the most stringent constraint, while the bounds from helium and deuterium are in good agreement with this range. No consistent solution is obtained from the lithium abundance, which reflects lithium problem. Nevertheless, since the model successfully satisfies the constraints from helium and deuterium, it can be considered viable from the perspective of Big Bang Nucleosynthesis. Thus, we establish the BBN viability of the model within a tightly constrained range of the parameter $n$, where our analysis yields significantly stronger bounds compared to those reported in the existing literature \cite{ref90,xx}, while remaining in good agreement with the results of other studies \cite{ref90,xx}.

\begin{table*}[htbp]
\centering
\caption{BBN constraints on model parameters.}
\resizebox{\textwidth}{!}{
\begin{tabular}{c|c|c|c|c|c}
\hline
\textbf{Model} & \textbf{Freeze-out} & \textbf{Helium} & \textbf{Deuterium} & \textbf{Lithium} & \textbf{Final Allowed Region} \\
\hline

Model I 
& $0 < b < 8.26 \times 10^{-56}$ 
& $0 < b < 3.65 \times 10^{-52}$ 
& $0 < b < 2.48 \times 10^{-51}$ 
& $6.6 \times 10^{-51} < b < 8.6 \times 10^{-51}$ 
& $0 < b < 8.26 \times 10^{-56}$ \\

\hline

Model II 
& $0 < b < 1.84 \times 10^{-109}$ 
& $0 < b < 6.35 \times 10^{-103}$ 
& $\begin{cases}
0 < b < 2.59 \times 10^{-101} \\
b > 1.41 \times 10^{-99}
\end{cases}$ 
& No allowed bound 
& $0 < b < 1.84 \times 10^{-109}$ \\

\hline

Model III 
& $0 < b < 3.77 \times 10^{-56}$ 
& $0 < b < 1.40 \times 10^{-52}$ 
& $0 < b < 7.46 \times 10^{-52}$ 
& $1.72 \times 10^{-51} < b < 2.19 \times 10^{-51}$ 
& $0 < b < 3.77 \times 10^{-56}$ \\

\hline

Model IV ($\gamma = 1$) 
& $0.999923 < n < 1.00008$ 
& $\begin{cases}
0.993856 < n < 1.00184 \\
1.02679 < n < 1.02818
\end{cases}$ 
& $\begin{cases}
0 < n < 1.00959 \\
1.02438 < n < 1.0303
\end{cases}$ 
& $1.03229 < n < 1.03304$ 
& $0.999923 < n < 1.00008$ \\

\hline
\end{tabular}
}
\label{t1}
\end{table*}

\section{Conclusion}
We explore in this work the viability of some cosmological models derived from modified entropic theories, by examining their consistency with Big Bang Nucleosynthesis (BBN) constraints. Since BBN provides one of the most stringent probes of the early Universe, any viable cosmological model must successfully reproduce the observed primordial abundances of light elements, particularly $^4\mathrm{He}$ and Deuterium.\\ 
Our analysis reveals that, across all considered models, the freeze-out condition imposes the most stringent constraint on the model parameters. The bounds obtained from helium and deuterium abundances are found to be mutually consistent and lie within the limits set by the freeze-out analysis. This consistency indicates that the models can successfully reproduce the observed abundances of $^4\mathrm{He}$ and Deuterium, thereby satisfying the primary requirements of BBN. On the other hand, the constraints obtained from lithium abundance do not overlap with those derived from helium and deuterium. This discrepancy reflects the well-known cosmological lithium problem, which persists even within the standard $\Lambda$CDM framework. Therefore, the inability to satisfy lithium constraints does not necessarily invalidate the models, as this issue remains unresolved in modern cosmology.\\
An important outcome of this study is that different models exhibit varying levels of constraint on their parameters. In particular, first three models \cite{k1,kruglov2025b,x} characterized by the parameter $b$ are typically constrained by upper bounds, often requiring extremely small deviations from General Relativity. In contrast, the forth model \cite{ref90} involving the parameter $n$ yield a narrow allowed interval, making them more predictive and tightly constrained. Furthermore, we have shown that the parameter values required to explain the late-time accelerated expansion of the Universe are naturally consistent with the bounds imposed by BBN. In aforementioned cases, the late-time estimates of the model parameters are found to lie well within the allowed BBN range, demonstrating that the models provide a unified description of both early- and late-time cosmological evolution.\\
In summary, the present analysis establishes that the considered modified cosmological models are viable from the perspective of BBN, provided their parameters lie within the constrained ranges obtained from freeze-out, helium, and deuterium observations. These results highlight the robustness of BBN as a powerful tool for testing cosmological models and emphasize the importance of combining early-Universe constraints with late-time observations to achieve a comprehensive understanding of cosmic evolution. A more thorough statistical analysis utilising Markov Chain Monte Carlo (MCMC) techniques may be used in future work to investigate potential parameter degeneracies and obtain strong constraints on the model parameters. Further tightening the boundaries and testing the model's consistency across many cosmological probes will require a thorough confrontation with a broader range of observational data sets, such as the Cosmic Microwave Background (CMB), Baryon Acoustic Oscillations (BAO), and Type Ia Supernovae. Additionally, it would be beneficial to expand the analysis to incorporate data on growth rates, structure development, and gravitational wave standard sirens, which can offer independent restrictions on the expansion history. Additionally, examining the model's dynamical behaviour using stability criteria and phase space analysis may provide more insight into its late-time evolution and physical feasibility.


\begin{thebibliography}{99}




\bibitem{bardeen1973}
J.~M.~Bardeen, B.~Carter and S.~W.~Hawking,
The four laws of black hole mechanics,
\textit{Commun. Math. Phys.} \textbf{31}, 161 (1973).
\href{https://doi.org/10.1007/BF01645742}{https://doi.org/10.1007/BF01645742}

\bibitem{bekenstein1973}
J.~D.~Bekenstein,
Black holes and entropy,
\textit{Phys. Rev. D} \textbf{7}, 2333 (1973).
\href{https://doi.org/10.1103/PhysRevD.7.2333}{https://doi.org/10.1103/PhysRevD.7.2333}

\bibitem{hawking1975}
S.~W.~Hawking,
Particle creation by black holes,
\textit{Commun. Math. Phys.} \textbf{43}, 199 (1975).
\href{https://doi.org/10.1007/BF02345020}{https://doi.org/10.1007/BF02345020}

\bibitem{tHooft1993}
G.~'t~Hooft,
Dimensional reduction in quantum gravity,
in \textit{Salamfestschrift}, World Scientific (1993).
\href{https://arxiv.org/abs/gr-qc/9310026}{https://arxiv.org/abs/gr-qc/9310026}

\bibitem{susskind1995}
L.~Susskind,
The world as a hologram,
\textit{J. Math. Phys.} \textbf{36}, 6377 (1995).
\href{https://doi.org/10.1063/1.531249}{https://doi.org/10.1063/1.531249}

\bibitem{maldacena1999}
J.~M.~Maldacena,
The large $N$ limit of superconformal field theories and supergravity,
\textit{Adv. Theor. Math. Phys.} \textbf{2}, 231 (1998).
\href{https://doi.org/10.1023/A:1026654312961}{https://doi.org/10.1023/A:1026654312961}

\bibitem{almheiri2013}
A.~Almheiri, D.~Marolf, J.~Polchinski and J.~Sully,
Black holes: Complementarity or firewalls?,
\textit{JHEP} \textbf{02}, 062 (2013).
\href{https://doi.org/10.1007/JHEP02(2013)062}{https://doi.org/10.1007/JHEP02(2013)062}

\bibitem{jacobson1995}
T.~Jacobson,
Thermodynamics of spacetime: The Einstein equation of state,
\textit{Phys. Rev. Lett.} \textbf{75}, 1260 (1995).
\href{https://doi.org/10.1103/PhysRevLett.75.1260}{https://doi.org/10.1103/PhysRevLett.75.1260}

\bibitem{padmanabhan2005}
T.~Padmanabhan,
Gravity and the thermodynamics of horizons,
\textit{Phys. Rept.} \textbf{406}, 49 (2005).
\href{https://doi.org/10.1016/j.physrep.2004.10.003}{https://doi.org/10.1016/j.physrep.2004.10.003}

\bibitem{verlinde2011}
E.~P.~Verlinde,
On the origin of gravity and the laws of Newton,
\textit{JHEP} \textbf{04}, 029 (2011).
\href{https://doi.org/10.1007/JHEP04(2011)029}{https://doi.org/10.1007/JHEP04(2011)029}

\bibitem{cai2005}
R.~G.~Cai and S.~P.~Kim,
First law of thermodynamics and Friedmann equations of FRW universe,
\textit{JHEP} \textbf{02}, 050 (2005).
\href{https://doi.org/10.1088/1126-6708/2005/02/050}{https://doi.org/10.1088/1126-6708/2005/02/050}

\bibitem{akbar2006}
M.~Akbar and R.~G.~Cai,
Thermodynamic behavior of Friedmann equations at apparent horizon,
\textit{Phys. Rev. D} \textbf{75}, 084003 (2007).
\href{https://doi.org/10.1103/PhysRevD.75.084003}{https://doi.org/10.1103/PhysRevD.75.084003}

\bibitem{sheykhi2018}
A.~Sheykhi,
Modified Friedmann equations from generalized entropy,
\textit{Phys. Lett. B} \textbf{785}, 118 (2018).
\href{https://doi.org/10.1016/j.physletb.2018.08.018}{https://doi.org/10.1016/j.physletb.2018.08.018}

\bibitem{nojiri2017}
S.~Nojiri, S.~D.~Odintsov and V.~K.~Oikonomou,
Modified gravity theories on a nutshell,
\textit{Phys. Rept.} \textbf{692}, 1 (2017).
\href{https://doi.org/10.1016/j.physrep.2017.06.001}{https://doi.org/10.1016/j.physrep.2017.06.001}

\bibitem{tsallis1988}
C.~Tsallis,
Possible generalization of Boltzmann--Gibbs statistics,
\textit{J. Stat. Phys.} \textbf{52}, 479 (1988).
\href{https://doi.org/10.1007/BF01016429}{https://doi.org/10.1007/BF01016429}

\bibitem{renyi1961}
A.~Rényi,
On measures of entropy and information,
in \textit{Proceedings of the Fourth Berkeley Symposium on Mathematical Statistics and Probability} (1961).

\bibitem{barrow2020}
J.~D.~Barrow,
The area of a rough black hole,
\textit{Phys. Lett. B} \textbf{808}, 135643 (2020).
\href{https://doi.org/10.1016/j.physletb.2020.135643}{https://doi.org/10.1016/j.physletb.2020.135643}

\bibitem{kaniadakis2002}
G.~Kaniadakis,
Statistical mechanics in the context of special relativity,
\textit{Phys. Rev. E} \textbf{66}, 056125 (2002).
\href{https://doi.org/10.1103/PhysRevE.66.056125}{https://doi.org/10.1103/PhysRevE.66.056125}

\bibitem{sharma1975}
B.~D.~Sharma and D.~P.~Mittal,
New non-additive measures of entropy,
\textit{J. Math. Sci.} \textbf{10}, 28 (1975).

\bibitem{rovelli1996}
C.~Rovelli,
Black hole entropy from loop quantum gravity,
\textit{Phys. Rev. Lett.} \textbf{77}, 3288 (1996).
\href{https://doi.org/10.1103/PhysRevLett.77.3288}{https://doi.org/10.1103/PhysRevLett.77.3288}


\bibitem{kruglov2025a}
Kruglov, S. I. (2025). Cosmology Due to Thermodynamics of Apparent Horizon. Annalen Der Physik. \href{https://doi.org/10.1002/andp.202500204}{doi:10.1002/andp.202500204}


\bibitem{kruglov2025b}
Kruglov, S. I. (2025). Cosmology, new entropy and thermodynamics of apparent horizon. Chinese Journal of Physics, 98, 277–286. \href{https://doi.org/10.1016/j.cjph.2025.08.045}{doi:10.1016/j.cjph.2025.08.045}


\bibitem{jahromi2018}
Sayahian Jahromi, A., Moosavi, S. A., Moradpour, H., Morais Graça, J. P., Lobo, I. P., Salako, I. G., \& Jawad, A. (2018). Generalized entropy formalism and a new holographic dark energy model. Physics Letters B, 780, 21–24. \href{https://doi.org/10.1016/j.physletb.2018.02.052}{doi:10.1016/j.physletb.2018.02.052}

‌
\bibitem{ren2021}
Ren, J. (2021). Analytic critical points of charged Rényi entropies from hyperbolic black holes. Journal of High Energy Physics, 2021(5). \href{https://doi.org/10.1007/jhep05(2021)080}{doi:10.1007/jhep05(2021)080}


\bibitem{mejrhit2019}
Mejrhit, K., \& Ennadifi, S-E. (2019). Thermodynamics, stability and Hawking–Page transition of black holes from non-extensive statistical mechanics in quantum geometry. Physics Letters B, 794, 45–49. \href{https://doi.org/10.1016/j.physletb.2019.03.055}{doi:10.1016/j.physletb.2019.03.055}

‌
\bibitem{majhi2017}
Majhi, A. (2017). Non-extensive statistical mechanics and black hole entropy from quantum geometry. Physics Letters B, 775, 32–36. \href{https://doi.org/10.1016/j.physletb.2017.10.043}{doi:10.1016/j.physletb.2017.10.043}


\bibitem{sekhmani2024}
Sekhmani, Y., Luciano, G. G., Maurya, S. K., Rayimbaev, J., Pourhassan, B., Jasim, M. K., \& Rincon, A. (2024). Exploring Tsallis thermodynamics for boundary conformal field theories in gauge gravity duality. Chinese Journal of Physics, 92, 894–914. \href{https://doi.org/10.1016/j.cjph.2024.10.015}{doi:10.1016/j.cjph.2024.10.015}


\bibitem{noorigashti2025}
Gashti, S. N., Pourhassan, B., Sakallı, İ., \& Brzo, A. B. (2025). Thermodynamic topology and photon spheres of dirty black holes within non-extensive entropy. Physics of the Dark Universe, 47, 101833. \href{https://doi.org/10.1016/j.dark.2025.101833}{doi:10.1016/j.dark.2025.101833}


\bibitem{sadeghi2014}
Sadeghi, J., Pourhassan, B., \& Abbaspour Moghaddam, Z. (2013). Interacting Entropy-Corrected Holographic Dark Energy and IR Cut-Off Length. International Journal of Theoretical Physics, 53(1), 125–135. \href{https://doi.org/10.1007/s10773-013-1790-1}{doi:10.1007/s10773-013-1790-1}

\bibitem{pourhassan2018}
Pourhassan, B., Bonilla, A., Faizal, M., \& Abreu, E. M. C. (2018). Holographic dark energy from fluid/gravity duality constraint by cosmological observations. Physics of the Dark Universe, 20, 41–48. \href{https://doi.org/10.1016/j.dark.2018.02.006}{doi:10.1016/j.dark.2018.02.006}

\bibitem{bbn11}Hubble, E. (1929). A relation between distance and radial velocity among extra-galactic nebulae. Proceedings of the National Academy of Sciences, 15(3), 168–173. \href{https://doi.org/10.1073/pnas.15.3.168}{doi:10.1073/pnas.15.3.168}

\bibitem{bbn12} Bethe, H. A. (1939). Energy Production in Stars. Physical Review, 55(5), 434–456. \href{https://doi.org/10.1103/physrev.55.434}{doi:10.1103/physrev.55.434}


\bibitem{bbn13} Capozziello, S., Lambiase, G., \& Saridakis, E. N. (2017). Constraining f(T) teleparallel gravity by big bang nucleosynthesis. The European Physical Journal C, 77(9). \href{https://doi.org/10.1140/epjc/s10052-017-5143-8}{doi:10.1140/epjc/s10052-017-5143-8}


\bibitem{bbn14} Jizba, P., \& Lambiase, G. (2023). Constraints on Tsallis Cosmology from Big Bang Nucleosynthesis and the Relic Abundance of Cold Dark Matter Particles. Entropy, 25(11), 1495. \href{https://doi.org/10.3390/e25111495}{doi:10.3390/e25111495}

\bibitem{bbna1} Asimakis, P., Basilakos, S., Mavromatos, N. E., \& Saridakis, E. N. (2022). Big bang nucleosynthesis constraints on higher-order modified gravities. Physical Review D, 105(8). \href{https://doi.org/10.1103/physrevd.105.084010}{doi:10.1103/physrevd.105.084010}

‌
\bibitem{bbna2} Torres, D. F., Vucetich, H., \& Plastino, A. (1997). Early Universe Test of Nonextensive Statistics. Physical Review Letters, 79(9), 1588–1590. \href{https://doi.org/10.1103/physrevlett.79.1588}{doi:10.1103/physrevlett.79.1588}


\bibitem{bbna3} Lambiase, G. (2005). Lorentz invariance breakdown and constraints from big-bang nucleosynthesis. Physical Review D, 72(8). \href{https://doi.org/10.1103/physrevd.72.087702}{doi:10.1103/physrevd.72.087702}


\bibitem{bbna4}Lambiase, G. (2012). Constraints on massive gravity theory from big bang nucleosynthesis. Journal of Cosmology and Astroparticle Physics, 2012(10), 028–028. \href{https://doi.org/10.1088/1475-7516/2012/10/028}{doi:10.1088/1475-7516/2012/10/028}


\bibitem{bbna5}Lambiase, G. (2011). Dark matter relic abundance and big bang nucleosynthesis in Horava’s gravity. Physical Review D, 83(10). \href{https://doi.org/10.1103/physrevd.83.107501}{doi:10.1103/physrevd.83.107501}

‌
\bibitem{bbn54}Barrow, J. D., Basilakos, S., \& Saridakis, E. N. (2021). Big Bang Nucleosynthesis constraints on Barrow entropy. Physics Letters B, 815, 136134. \href{https://doi.org/10.1016/j.physletb.2021.136134}{doi:10.1016/j.physletb.2021.136134}


\bibitem{s1} Daniel, C., Pereira, D. S., Francisco, \& Mimoso, José P. (2025). Big Bang Nucleosynthesis constraints on $f(T,L_m)$ gravity. \href{https://arxiv.org/abs/2509.20309}{arxiv.org/abs/2509.20309}

\bibitem{s2} Sheykhi, A., Sooraki, A. S., \& Liravi, L. (2025). Big bang nucleosynthesis constraints on dual Kaniadakis cosmology. Physical Review D, 112(10). \href{https://doi.org/10.1103/fg96-fjnw}{doi:10.1103/fg96-fjnw}

\bibitem{s3} Ge, J., Ming, L., Liang, S.-D., Zhang, H.-H., \& Harko, T. (2024). Constraining Weyl type f(Q,T) gravity with Big Bang Nucleosynthesis. \href{https://arxiv.org/abs/2407.10421}{arxiv.org/abs/2407.10421}

\bibitem{r1} Padmanabhan, T. (2005). Gravity and the thermodynamics of horizons. Physics Reports, 406(2), 49–125. \href{https://doi.org/10.1016/j.physrep.2004.10.003}{https://doi.org/10.1016/j.physrep.2004.10.003}

\bibitem{r2} Frolov, A. V., \& Kofman, L. (2003). Inflation and de Sitter thermodynamics. Journal of Cosmology and Astroparticle Physics, 2003(05), 009–009. \href{https://doi.org/10.1088/1475-7516/2003/05/009}{https://doi.org/10.1088/1475-7516/2003/05/009}

\bibitem{r3} Gibbons, G. W., \& Hawking, S. W. (1977). Cosmological event horizons, thermodynamics, and particle creation. Physical Review D, 15(10), 2738–2751. \href{https://doi.org/10.1103/physrevd.15.2738}{https://doi.org/10.1103/physrevd.15.2738}



\bibitem{k1}I, K. S. (2025). New entropy, thermodynamics of apparent horizon and cosmology. Retrieved January 8, 2026, from arXiv.org website: \href{https://arxiv.org/abs/2502.12165‌}{https://arxiv.org/abs/2502.12165‌}

\bibitem{b} Barrow, J. D. (2020). The area of a rough black hole. Physics Letters B, 808, 135643–135643. \href{https://doi.org/10.1016/j.physletb.2020.135643}{https://doi.org/10.1016/j.physletb.2020.13564}

\bibitem{t} Caruso, F., \& Tsallis, C. (2008). Nonadditive entropy reconciles the area law in quantum systems with classical thermodynamics. Physical Review E, 78(2). https://doi.org/10.1103/physreve.78.021102

‌

\bibitem{k3} Mohammadi, H., \& Salehi, A. (2023). Friedmann equations with the generalized logarithmic modification of Barrow entropy and Tsallis entropy. Physics Letters B, 839, 137794–137794. \href{https://doi.org/10.1016/j.physletb.2023.137794}{https://doi.org/10.1016/j.physletb.2023.137794}


\bibitem{x} Kruglov,  S. Entropic Cosmology with New Entropy. Preprints 2026, 2026030027. \href{https://doi.org/10.20944/preprints202603.0027.v1}{doi:10.20944/preprints202603.0027.v1}


\bibitem{ref90} Basilakos, S., Lymperis, A., Petronikolou, M., \& Saridakis, E. N. (2025). Modified cosmology through spacetime thermodynamics and generalized mass-to-horizon entropy. The European Physical Journal C, 85(11). \href{https://doi.org/10.1140/epjc/s10052-025-14971-8}{doi:10.1140/epjc/s10052-025-14971-8}


\bibitem{70} K. A. Olive, G. Steigman, T. P. Walker, 
Primordial nucleosynthesis: theory and observations, 
\textit{Phys. Rep.} \textbf{333}, 389 (2000).

\bibitem{71} R. H. Cyburt et al., 
Big bang nucleosynthesis: present status, 
\textit{Phys. Rev. Mod. Phys.} \textbf{88}, 015004 (2016). 
\href{https://doi.org/10.1103/PhysRevModPhys.88.015004}{https://doi.org/10.1103/PhysRevModPhys.88.015004}

\bibitem{72} D. F. Torres, H. Vucetich, A. Plastino, 
Early universe test of non-extensive statistics, 
\textit{Phys. Rev. Lett.} \textbf{79}, 1588 (1997). 
\href{https://doi.org/10.1103/PhysRevLett.79.1588}{https://doi.org/10.1103/PhysRevLett.79.1588}

\bibitem{73} G. Lambiase, 
Dark matter relic abundance and big bang nucleosynthesis in Hořava’s gravity, 
\textit{Phys. Rev. D} \textbf{83}, 107501 (2011). 
\href{https://doi.org/10.1103/PhysRevD.83.107501}{https://doi.org/10.1103/PhysRevD.83.107501}

\bibitem{74} K. A. Olive et al., 
Review of particle physics, 
\textit{Chin. Phys. C} \textbf{38}, 0900001 (2014). 
\href{https://doi.org/10.1088/1674-1137/38/9/090001}{https://doi.org/10.1088/1674-1137/38/9/090001}

\bibitem{75} G. Lambiase, 
Lorentz invariance breakdown and constraints from big-bang nucleosynthesis, 
\textit{Phys. Rev. D} \textbf{72}, 087702 (2005). 
\href{https://doi.org/10.1103/PhysRevD.72.087702}{https://doi.org/10.1103/PhysRevD.72.087702}

\bibitem{76} K. A. Olive, E. Skillman, G. Steigman, 
The primordial abundance of $^4$He: an update, 
\textit{Astrophys. J.} \textbf{483}, 788 (1997). 
\href{https://doi.org/10.1086/304265}{https://doi.org/10.1086/304265}


\bibitem{27}Scott Dodelson. Modern Cosmology. Academic Press,
Amsterdam, 2003. ISBN 978-0-12-219141-1.

\bibitem{4} Steigman, G. (2012). Neutrinos and Big Bang Nucleosynthesis. Advances in High Energy Physics, 2012, 1–24. \href{https://doi.org/10.1155/2012/268321}{doi:10.1155/2012/268321}

\bibitem{28} E. Komatsu et al. Seven-yearwilkinson microwave
anisotropy probe(wmap) observations: Cosmological interpretation. The Astrophysical Journal Supplement Series, 192(2):18, January 2011. ISSN 1538-4365. href{https://doi.org/10.1088/0067-0049/192/2/18}{doi:
10.1088/0067-0049/192/2/18.}‌

\bibitem{29} Fields, B. D., Olive, K. A., Yeh, T.-H., \& Young, C. (2020). Erratum: Big-Bang Nucleosynthesis after Planck. Journal of Cosmology and Astroparticle Physics, 2020(11), E02–E02. \href{https://doi.org/10.1088/1475-7516/2020/11/e02}{doi:10.1088/1475-7516/2020/11/e02}

\bibitem{31} Steigman, G. (2007). Primordial Nucleosynthesis in the Precision Cosmology Era. Annual Review of Nuclear and Particle Science, 57(1), 463–491. \href{https://doi.org/10.1146/annurev.nucl.56.080805.140437}{doi:10.1146/annurev.nucl.56.080805.140437}


‌\bibitem{57} Boran, S., \& Kahya, E. O. (2014). Testing a Dilaton Gravity Model Using Nucleosynthesis. Advances in High Energy Physics, 2014, 1–7. \href{https://doi.org/10.1155/2014/282675}{doi:10.1155/2014/282675}

‌\bibitem{xx} Luciano, G. G., \& Saridakis, E. N. (2025). Baryogenesis constraints on generalized mass-to-horizon entropy.\href{ https://arxiv.org/abs/2511.01693}{Arxiv:2511.01693}

\bibitem{xy} Luciano, G. G., \& Paliathanasis, A. (2025). Modified cosmology through generalized mass-to-horizon entropy: Observational constraints from DESI DR2 BAO data. Physics Letters B, 870, 139954. \href{https://doi.org/10.1016/j.physletb.2025.139954}{doi:10.1016/j.physletb.2025.139954}
‌
‌

‌



\end{thebibliography}
\end{document}